\documentclass[10pt,aps,prl,twocolumn,superscriptaddress]{revtex4-2}

\usepackage{color}
\usepackage{amsmath}
\usepackage{graphicx}
\usepackage{slashed}
 \usepackage{ulem}
\usepackage{hyperref}
\hypersetup{
    colorlinks=true,
    linkcolor=blue,
    filecolor=magenta,      
    urlcolor=cyan,
    citecolor=blue
}

\begin{document}

\title{An axion constraint from the diffuse supernova neutrino background indicated by Super-Kamiokande}

\author{Kanji Mori}
\email[]{kanji.mori@rk.phys.keio.ac.jp}
\affiliation{Department of Physics, Faculty of Science and Technology, Keio University, 3-14-1 Hiyoshi, Kohoku-ku, Yokohama, Kanagawa 223-8522 Japan}
\author{Tomoya Takiwaki}
\affiliation{National Astronomical Observatory of Japan, 2-21-1 Osawa, Mitaka, Tokyo 181-8588, Japan}
\affiliation{The Graduate University for Advanced Studies (SOKENDAI), 2-21-1 Osawa, Mitaka, Tokyo 181-8588, Japan}
\author{Kazunori Kohri}
\affiliation{National Astronomical Observatory of Japan, 2-21-1 Osawa, Mitaka, Tokyo 181-8588, Japan}
\affiliation{Department of Astronomy, The University of Tokyo, Bunkyo-ku, Hongo, Tokyo 113-0033, Japan}
\affiliation{The Graduate University for Advanced Studies (SOKENDAI), 2-21-1 Osawa, Mitaka, Tokyo 181-8588, Japan}
\affiliation{Theory Center, IPNS, KEK, 1-1 Oho, Tsukuba, Ibaraki 305-0801, Japan}
\affiliation{Kavli IPMU (WPI), The University of Tokyo, 5-1-5 Kashiwanoha, Kashiwa, Chiba 277-8583, Japan}
\author{Masamitsu Mori}
\affiliation{National Institute of Technology (KOSEN), Numazu College, 3600 Ooka, Numazu, Shizuoka 410-0022, Japan}

\date{\today}

\begin{abstract}

Recently, the Super-Kamiokande Collaboration reported an indication of the diffuse supernova neutrino background (DSNB) with a statistical significance of $2.6\sigma$. Motivated by this possible discovery, we investigate the impact of axion cooling on the DSNB flux on the basis of long-term neutrino-radiation hydrodynamic simulations. We compare the observed flux and our models and obtain a $1\sigma$ upper limit $|g_{ap}|<1.3\times10^{-9}$ on the axion-proton coupling constant, which is comparable to the conventional limit based on the SN~1987A neutrino burst. In contrast to the SN~1987A bound, the DSNB constraint does not rely on the properties of a single observed supernova, because the DSNB represents the cumulative neutrino emission from a cosmic population of core-collapse events. More generally, this approach can be applied to other feebly interacting particles that modify protoneutron-star cooling.
\end{abstract}

\maketitle

\section{Introduction}

Core-collapse supernovae are powerful laboratories for particles and interactions beyond the Standard Model \cite[e.g.,][]{Raffelt1990}. Unlike terrestrial experiments, they probe new physics through its impact on matter at extreme temperature and density, most notably through anomalous energy loss from the newly formed protoneutron star. 
Recent advances in neutrino-radiation hydrodynamic simulations~\citep[e.g.,][]{2024ApJ...964L..16B,Janka2025,2025MNRAS.536..280N} have made increasingly detailed predictions of the neutrino emission from core-collapse supernovae possible, providing a firmer theoretical basis for such tests of new physics. Neutrino observations therefore offer a particularly direct means of probing exotic particles that can modify the cooling of the protoneutron star.

The neutrino detection from SN~1987A provided the first observational opportunity to exploit this idea~\cite{1987PhRvL..58.1490H,1987PhRvL..58.1494B,1987JETPL..45..589A}.
The detection established core-collapse supernovae as intense neutrino sources, with an emitted neutrino energy of order $10^{53}$\,erg. The observed duration and energetics of the burst have since provided stringent constraints on exotic energy-loss mechanisms.
However, Galactic supernovae are estimated to be as rare as $\sim3$ events per century \cite{2025A&A...698A.306M}. 

Instead of searching for neutrinos from a single supernova event, the idea of the diffuse supernova neutrino background (DSNB) has been discussed \cite{1984Natur.310..191K,1984NYASA.422..319B,1986ApJ...302...19W,1995APh.....3..367T,1997APh.....7..125M,HARTMANN1997137,2006APh....26..190L,2009PhRvL.102w1101L,2009PhRvD..79h3013H,2012JCAP...07..012L,2014ApJ...790..115M,2015ApJ...804...75N,2017JCAP...11..031P,2018JCAP...05..066M,2018MNRAS.475.1363H,2021ApJ...909..169K,2022ApJ...937...30A,2023ApJ...953..151A,2024ApJ...975...71N,2024PhRvD.110j3029M,2024PhRvD.109b3024E,2026PhRvD.113f3044L}, which is a steady neutrino flux originating from many core-collapsing stars throughout the history of the Universe. Because it integrates neutrino emission over a large population of core-collapse events, the DSNB carries information on both supernova physics and the cosmic core-collapse population. This has motivated extensive searches by several neutrino experiments, including SNO, Borexino, KamLAND, and Super-Kamiokande~\cite{2004PhRvD..70i3014A,AGOSTINI2021102509,2026ApJ..1005..101A,2026arXiv260629381C}. Recently, at Neutrino 2026: XXXII International Conference on Neutrino Physics and Astrophysics, the SK Collaboration reported the first indication of DSNB with a statistical significance of $2.6\sigma$ on the basis of $\sim5000$ days of its dataset \cite{SKreport}. Although the significance is not high enough for the clear detection, the signal may be the second discovery of supernova neutrinos, following the historic SN~1987A event.

Motivated by the recent indication of the DSNB, we investigate its potential as a probe of physics beyond the Standard Model.
The DSNB has previously been studied as a probe of new physics, including sterile neutrinos~\cite{2018JCAP...06..019J}, neutrino decay~\cite{2004PhRvD..70a3001F,2021JCAP...05..011T,2023PhRvD.107b3017I,2026JCAP...02..069I}, and non-standard neutrino interactions~\cite{2006JHEP...11..023G,2007PhRvD..76f3004B,2021PhRvD.103b3527C,2022PhRvD.106i5042D,2022JHEP...12..050A,2025PhRvL.135r1002W,2025JCAP...01..062M,2026arXiv260622898B}. Here, we focus on a different class of effects: additional cooling of the supernova core induced by axion emission.
Axions were originally introduced as a solution to the strong CP problem~\cite{1978PhRvL..40..279W,1978PhRvL..40..223W} and are also well-motivated dark-matter candidates~\cite{1983PhLB..120..133A,1983PhLB..120..127P,1983PhLB..120..137D}. They have consequently been the subject of extensive laboratory and astrophysical searches~\cite[e.g.][]{2021ARNPS..71..225C,Caputo:2024PM,2025PhyR.1117....1C}.

Among astrophysical environments, core-collapse supernovae can be a copious source of axions. For sufficiently weak couplings, axions produced in the hot and dense core can escape  freely and provide an additional energy-loss channel, accelerating PNS cooling and reducing the neutrino emission. In fact, the neutrino burst detected from SN~1987A has been used to constrain the axion mass and the axion-nucleon coupling constant \cite{1988PhRvL..60.1797T,PhysRevD.39.1020,1988PhRvL..60.1793R,1990PhRvL..65..960E,1997PhRvD..56.2419K,2019JCAP...10..016C,2022PhRvD.105c5022C,2024PhRvD.109b3001L}. However, any constraint inferred from SN~1987A necessarily relies on a single supernova event and may therefore be sensitive to the particular properties of its progenitor and explosion. For example, Ref.~\cite{2020PhRvD.101l3025B} pointed out that the traditional SN~1987A limit can be evaded if the stellar core rapidly rotates and an accretion disk around a central black hole is formed. 

The DSNB naturally mitigates this single-event limitation because it represents the cumulative neutrino emission from a large population of core-collapse supernovae. In fact, the idea of the diffuse \textit{axion} supernova background (DSAB) has been also pursued \cite{2011PhRvD..84j3008R}, which is an axion analogue of DSNB. Detecting DSAB is much more challenging than detecting DSNB because of the extremely feeble interactions. Still,  $\gamma$-ray production from the axion-photon conversion in the interstellar magnetic field has been used to obtain constraints on axionlike particles \cite{2020PhRvD.102l3005C,2022PhRvD.105f3028C,2025PhRvD.112a5006C} which do not rely on SN~1987A. 

   \begin{figure}
        \centering
        \includegraphics[width=\hsize]{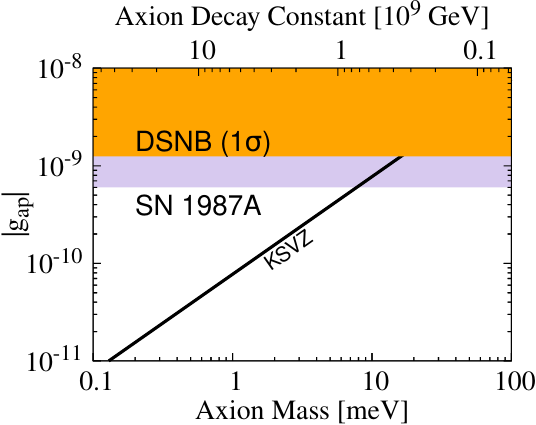}
        \caption{Upper bounds on the axion-proton coupling constant $g_{ap}$ as a function of axion mass in meV. The orange region is excluded by the DSNB.
        The purple region shows the conventional SN~1987A cooling constraint based on the Raffelt criterion~\cite{2024PhRvD.109b3001L}.
        }\label{gap}%
    \end{figure}

 Our result is summarized in Fig.~\ref{gap}. Assuming Kim-Shifman-Vainshtein-Zakharov (KSVZ) axions \cite{1979PhRvL..43..103K,1980NuPhB.166..493S}, we obtain an upper limit
 \begin{equation}
     |g_{ap}|<1.3\times10^{-9}\quad(1\sigma)\label{lim_g}
 \end{equation}
 on the axion-proton coupling constant $g_{ap}$ by comparing the SK observation and our DSNB models.
For comparison, Fig.~\ref{gap} also shows the conventional SN~1987A cooling bound based on the Raffelt criterion; its connection to the PNS cooling evolution is discussed in Fig.~\ref{nub_L}.
 The new limit is comparable to the SN~1987A constraint, $|g_{ap}|<6\times10^{-10}$~\cite{2024PhRvD.109b3001L} and is free from the bias that originates from the single supernova event.
 
 In the KSVZ model, the coupling constant is given by $g_{ap}=-0.47m_\mathrm{p}/f_a$ \cite[e.g.][]{2016JHEP...01..034D}, where $m_p$ is the proton mass and $f_a$ is the axion decay constant. In terms of $f_a$, the limit in Eq.~(\ref{lim_g}) corresponds to
  \begin{equation}
f_a>0.35\times10^9\,\mathrm{GeV}.
 \end{equation}
 The axion mass $m_a$ can be determined by a relation $m_a=5.7(10^9\,\mathrm{GeV}/f_a)\,\mathrm{meV}$, leading to a limit 
 \begin{equation}
     m_a<16\,\mathrm{meV}.
 \end{equation}

 \section{Supernova Simulations}
 \label{sec:sn}

   \begin{figure}
        \centering
        \includegraphics[width=\hsize]{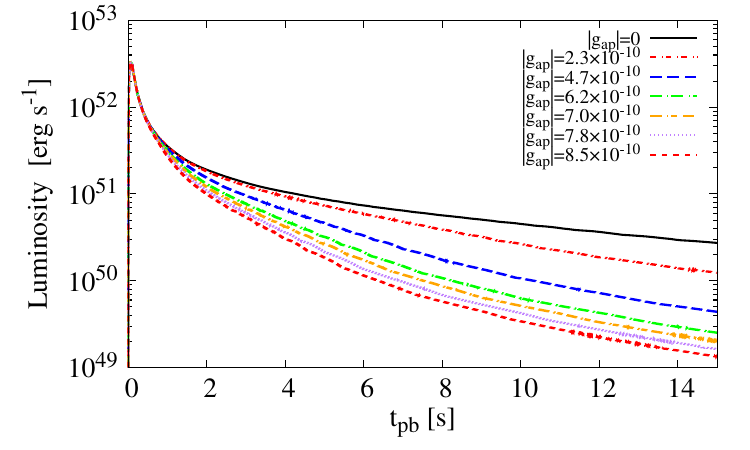}
        \caption{The $\bar{\nu}_e$ luminosity as a function of the post-bounce time $t_\mathrm{pb}$ for the supernova models in Ref.~\cite{2025JCAP...11..081M}. The solid curve indicates the model without axions and the dashed curves indicate the models with axions with different values for $|g_{ap}|$.}
        \label{nub_L}%
    \end{figure}

In order to evaluate the effect of axions on DSNB, we use supernova models developed by \citet{2025JCAP...11..081M}, who performed one-dimensional general-relativistic neutrino-radiation hydrodynamic simulations coupled with axion cooling that last for $\sim15$\,s after the core bounce. The simulations adopted a $9.6M_\odot$ star with zero-metallicity as the progenitor model, which can lead to a successful explosion even in spherically-symmetric geometry.
The axion-proton interaction was described by
\begin{equation}
    \mathcal{L}=\frac{g_{ap}}{2m_p}\bar{p}\gamma^\mu\gamma_5p\partial_\mu a.
\end{equation}
As the axion production channel, the nucleon bremsstrahlung was taken into account with the one-pion-exchange (OPE) approximation \cite{1988PhRvD..38.2338B,1995PhRvD..52.1780R} and the nucleon spin fluctuation effect \cite{1997PhRvD..56.2419K}. 
Seven models with axion cooling were calculated with $m_a\leq11$\,meV. Assuming the KSVZ model, this corresponds to $g_{ap}\leq8.5\times10^{-10}$ and the axion-neutron coupling constant $g_{an}\sim0$. Since the coupling constants are tiny, axions freely escape from the star once they are produced in the protoneutron star (PNS). Hence they work as an additional cooling process and can reduce the neutrino luminosity of supernova events. 

Figure \ref{nub_L} shows the $\bar{\nu}_e$ light curves for the supernova models with and without axion cooling. In the model without axions, the $\bar{\nu}_e$ luminosity reaches the peak at $t_\mathrm{pb}\sim0.05$\,s and it starts decreasing, following the cooling of the PNS. If we consider the axion cooling, the $\bar{\nu}_e$  luminosity starts deviating from the standard case at $t_\mathrm{pb}\sim1$\,s, because the PNS cooling is accelerated by axions. This accelerated cooling shortens the neutrino-emission timescale and forms the basis of the conventional SN~1987A cooling bound on axions. A commonly used implementation is the Raffelt criterion, which requires the axion luminosity not to exceed the neutrino luminosity around $t_\mathrm{pb}=1$\,s.
 

 \section{DSNB}
 \label{sec:DSNB}

   \begin{figure}
        \centering
        \includegraphics[width=\hsize]{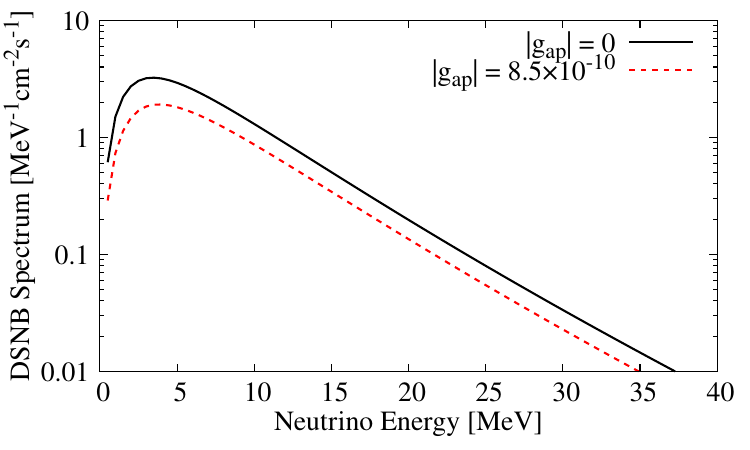}
        \caption{The DSNB spectrum $\Phi_{\bar{\nu}_e}(E)$ with $|g_{ap}|=0$ and $8.5\times10^{-10}$. The spectra for $|g_{ap}|<8.5\times10^{-10}$ fall between the two models.}
        \label{DSNB_spec}%
    \end{figure}

DSNB is the accumulation of supernova neutrinos throughout the cosmic history. If we neglect the neutrino oscillations, the DSNB $\bar{\nu}_e$ spectrum can be evaluated as~\cite[e.g.][]{2023PJAB...99..460A}

\begin{equation}
    \Phi_{\bar{\nu}_e}(E)=\frac{c}{H_0}\int^{z_\mathrm{max}}_0\frac{\dot\rho(z)f_{\bar{\nu}_e}(E')}{\sqrt{\Omega_\mathrm{m}(1+z)^3+\Omega_\mathrm{\Lambda}}}dz,\label{DSNB_eq}
\end{equation}
where $z$ is the redshift, $E$ is the neutrino energy, $E'=E(1+z)$, $H_0$ is the Hubble constant, $f_{\bar{\nu}_e}(E')$ is the neutrino spectrum of each supernova event averaged by the stellar population,  and $\Omega_\mathrm{m}$ and $\Omega_\Lambda$ are the cosmological matter and dark energy density, respectively. We reconstruct the neutrino spectrum with the pinched quasithermal distribution \cite{2003ApJ...590..971K}
\begin{equation}
    f_{\bar{\nu}_e}(E)=\int \frac{\dot{N}(1+\alpha)^{1+\alpha}}{\Gamma(1+\alpha)\langle E\rangle}\left(\frac{E}{\langle E\rangle}\right)^\alpha\exp\left(-(1+\alpha)\frac{E}{\langle E\rangle}\right)dt,
\end{equation}
where $\dot{N}$ is the number luminosity, $\langle E\rangle$ is the mean energy, and $\alpha$ is defined by an equation $(2+\alpha)/(1+\alpha)=\langle E_\mathrm{rms}\rangle/\langle E\rangle$ with $\langle E_\mathrm{rms}\rangle$ the root-mean-square energy. The stellar core-collapse event rate is estimated by an empirical formula
\begin{equation}
    \dot{\rho}(z)=\dot{\rho}_0\left((1+z)^{a\eta}+\left(\frac{1+z}{B}\right)^{b\eta}+\left(\frac{1+z}{C}\right)^{c\eta}\right)^\frac{1}{\eta},
\end{equation}
with the same values for  $\dot{\rho}_0,\,a,\,b,\,c,\,B,\,C,$ and $\eta$ as in Ref.~\cite{2026PhRvD.113f3044L}. We take $z_\mathrm{max}=4.5$ for the interval of integration, considering a decline in the supernova rate at $z>4$.

We adopt a DSNB model developed in Ref.~\cite{2026PhRvD.113f3044L} based on the ``KIHNN" binary population synthesis code \cite{2014MNRAS.442.2963K} as the fiducial spectrum without axions. As for the effect of axions, we first evaluate Eq.~(\ref{DSNB_eq}) with the neutrino spectrum of each supernova model in Ref.~\cite{2025JCAP...11..081M}. Then we calculate the reduction factor $p_{\bar{\nu}_e}(g_{ap},\,E)=\Phi_{\bar{\nu}_e}(g_{ap},\,E)/\Phi_{\bar{\nu}_e}(g_{ap}=0,\,E)$ and multiply it by the fiducial DSNB spectrum.

Fig.~\ref{DSNB_spec} shows the DNSB spectrum $\Phi_{\bar{\nu}_e}(E)$ for $|g_{ap}|=0$ and $8.5\times10^{-10}$. The results with $|g_{ap}|<8.5\times10^{-10}$ fall between the two models. One can find that DSNB is suppressed by the axion cooling. At $E=10$\,MeV, the reduction factor is estimated to be $p_{\bar{\nu}_e}(g_{ap}=8.5\times10^{-10},\,E=10\,\mathrm{MeV})=0.67$.

   \begin{figure}
        \centering
        \includegraphics[width=\hsize]{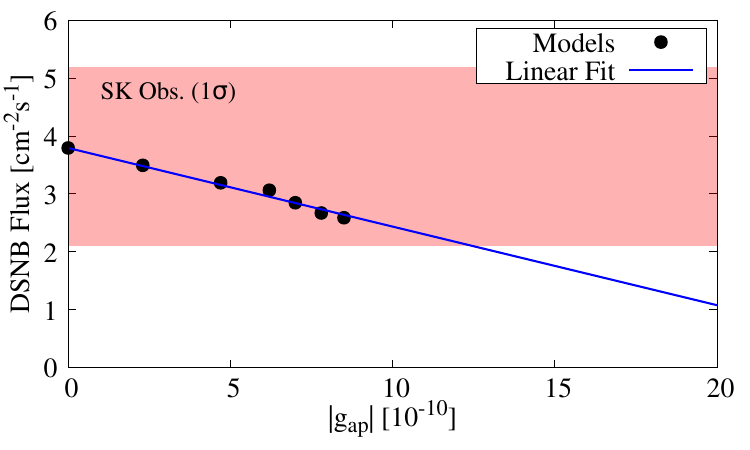}
        \caption{The DSNB flux $F_{\bar{\nu}_e}$ as a function of $|g_{ap}|$. The points indicate the flux estimated from the supernova models. The linear fit of the models is shown in the blue line. The red band shows the $1\sigma$ flux reported by SK \cite{SKreport}.}
        \label{flux}%
    \end{figure}

Recently, the SK Collaboration reported an indication of the DSNB flux $F_{\bar{\nu}_e}=3.6^{+1.6}_{-1.5}$\,cm$^{-2}$s$^{-1}$ for the neutrino energy $E=13.3$--81.3\,MeV \cite{SKreport}. In order to compare the observed DSNB flux with our models, we estimate the flux as
\begin{equation}
    F_{\bar{\nu}_e}(g_{ap})=\int^{E_\mathrm{max}}_{E_\mathrm{min}}\Phi_{\bar{\nu}_e}(g_{ap},\,E)dE.
\end{equation}
Considering the SK threshold, we fix the minimum energy threshold $E_\mathrm{min}=13.3$\,MeV. As for the maximum energy threshold, we adopt $E_\mathrm{max}=60$\,MeV because Ref.~\cite{2026PhRvD.113f3044L} does not provide the DSNB spectrum above this energy. This choice is justified because the contribution of neutrinos with $E=60$--81.3\,MeV to the flux is negligible.

Figure \ref{flux} shows the DSNB flux $F_{\bar{\nu}_e}(g_{ap})$ estimated from the supernova models. The flux decreases as the axion-proton coupling constant increases because the PNS is cooled down by the axion emission. Ref.~\cite{2025JCAP...11..081M} performed neutrino-radiation hydrodynamic simulations only for axions with $|g_{ap}|\leq8.5\times10^{-10}$. As the figure shows, the model flux is consistent with the observation in this parameter region. In order to estimate the DSNB flux for $|g_{ap}|>8.5\times10^{-10}$, we fit the model prediction with a linear function. The figure shows that the extrapolated prediction contradicts the observation for $|g_{ap}|>1.3\times10^{-9}$. This leads to the DSNB constraint on $g_{ap}$ shown in Fig.~\ref{gap}.


\section{Discussion}\label{sec:sum-dis}

In this Letter, we estimated the impact of axion cooling on the DSNB flux based on neutrino-radiation hydrodynamic simulations. Comparing the DSNB flux reported by SK with our models, we obtained a $1\sigma$ upper limit, $|g_{ap}|<1.3\times10^{-9}$, on the axion-proton coupling constant. 
This limit is comparable to the conventional one based on the SN~1987A neutrino burst. The two constraints, however, probe different aspects of supernova neutrino emission. The conventional SN~1987A cooling bound is primarily sensitive to the duration and temporal evolution of the neutrino burst, whereas the DSNB constraint obtained here probes the time-integrated neutrino emission. In addition, while the SN~1987A bound relies on a single supernova event, the DSNB averages over neutrino emission from a large population of core-collapse supernovae throughout cosmic history.

The axion emissivity from a PNS is subject to theoretical uncertainties. The simulations in Ref.~\cite{2025JCAP...11..081M} adopted the modified OPE approximation, taking into account the nucleon spin fluctuation \cite{1997PhRvD..56.2419K}. Recently, the treatment beyond the OPE approximation was discussed in the context of the PNS cooling \cite{2018JHEP...09..051C,2019JCAP...10..016C}. In addition, other axion production channels such as the pion-induced reaction~\cite{1995PhRvD..52.1780R,1997PhRvD..56.2419K,2021PhRvL.126g1102C,2021PhRvD.104j3012F} and processes involving strange matter \cite{2024PhRvL.133l1002C} are being investigated. It is desirable to perform self-consistent simulations coupled with these processes to render the cooling limits more accurate.

In this study, we adopted the $9.6M_\odot$ star as a progenitor model to estimate the reduction factor $p_{\bar{\nu}_e}(g_{ap},\,E)$ for the DSNB spectrum. However, DSNB is a superposition of supernova events with a wide range of mass. Since the $9.6M_\odot$ model has a tiny compactness parameter of $\xi_{2.5}=7.6\times10^{-5}$ \cite{2017ApJ...850...43R}, it is expected that the temperature in the PNS and hence the axion luminosity are lower than those in heavier stars. This suggests that our limit is conservative, but it is still important to perform simulations for a variety of stars to estimate uncertainties in the DSNB flux.

So far, we have not considered the effect of neutrino oscillations. The oscillations are often parametrized as \cite{2000PhRvD..62c3007D}
\begin{equation}
    \Phi_{\bar\nu_e}(E)
= P \Phi^0_{\bar\nu_e}(E)
+ (1 - P) \Phi^0_{\bar\nu_x}(E),\label{osci}
\end{equation}
where $\Phi^0_{\bar\nu_e}(E)$ and  $\Phi^0_{\bar\nu_x}(E)$ are the DSNB spectra  for $\bar\nu_e$ and $\bar\nu_x$, respectively, without the neutrino oscillations and $P\in[0,\,1]$ is the survival probability.  If we consider the vacuum oscillation in the interstellar space and the Mikheyev-Smirnov-Wolfenstein  effect in the stellar envelope, the survival probability is estimated as $P=\cos^2\theta_{12}\cos^2\theta_{13}$  for the normal mass hierarchy and $P=\sin^2\theta_{13}$ for the inverted hierarchy using the mixing angles $\theta_{12}$ and $\theta_{13}$. However, we adopted $P=1$ in this study because the effect of collective oscillations on $P$ induced by the neutrino self-interaction is highly uncertain. Motivated by this uncertainty, Ref.~\cite{2026PhRvD.113f3044L} investigated the oscillation effect by using Eq.~(\ref{osci}) and reported that the DSNB flux can be reduced by a few tens of percent if the most extreme case of $P=0$ is adopted. This suggests that our limit is conservative in terms of oscillations.  Still, the treatment of the collective oscillations in supernovae should be established to discuss the impact on DSNB, although it is still under debate \cite[e.g.][]{2025ARNPS..75..399J}.

This method based on DSNB is applicable not only to axions but also to a wide range of exotic feebly-interacting particles, including axionlike particles \cite{2020JCAP...12..008L,2022PhRvD.105c5022C,2023PhRvD.108f3027M,2024PhRvD.109b3001L,2026PhRvD.113f3045M,2026arXiv260417840T}, sterile neutrinos \cite{2011PhRvD..83i3014R,2020JCAP...01..010M,2024PhRvD.109f3010C,2024JHEP...07..057A,2024PhRvD.110b3031M}, and dark photons \cite{2016PhRvC..94d5805R,2017JHEP...01..107C,2026PhRvD.113d1303M}. It is desirable to discuss the impact of such particles on DSNB to establish a variety of constraints that do not depend on the single event of SN~1987A.

\begin{acknowledgments}
This work was in part supported by JSPS KAKENHI Grant Numbers, JP23K13107, JP25H02194 (K.M), JP23K25895, JP23K22494, JP23K03400, JP24K00631 (T.T.), 
and JP24K07027 (K.K.).
This work was also supported by “Program for Promoting researches on the Supercomputer Fugaku” (Structure and Evolution of the Universe Unraveled by Fusion of Simulation and AI; Grant Number JPMXP1020230406), and JICFuS.
\end{acknowledgments}

\bibliography{ref.bib}
\end{document}